\documentclass[12pt]{article}
\usepackage{amssymb}
\usepackage{amstext}
\usepackage{amsmath}
\usepackage{latexsym}
\usepackage{color}

\newcommand{\mb}{\mathbb}

\newcommand{\be}{\begin{equation}}
\newcommand{\en}{\end{equation}}

\newtheorem{thm}{Theorem}
\newtheorem{cor}[thm]{Corollary}

\newtheorem{prop}[thm]{Proposition}

\newtheorem{defi}{Definition}[section]
\newtheorem{lem}[defi]{Lemma}
\newtheorem{theorem}[defi]{Theorem}

\newcommand{\bedefin}{\begin{defi}}
\newcommand{\findefi}{\end{defi} \medskip}

\newcommand{\betheo}{\begin{theorem}$\!\!${\bf \,\,\,}}
\newcommand{\entheo}{\end{theorem}}
\newcommand{\enth}{\end{theorem}}

\newcommand{\becor}{\begin{cor}$\!\!${\bf .}}
\newcommand{\encor}{\end{cor}}

\newcommand{\belem}{\begin{lem}$\!\!${\bf .$\quad$}}
\newcommand{\enlem}{\end{lem}}

\newcommand{\prf}{\noindent{\bf{\small Proof.}\,\,\,\,}}
\newcommand{\qed}{\hfill $\blacksquare$}
\newcommand{\eg}{\noindent{\bf{\small Example}\,}}

\newcommand{\bea}{\begin{eqnarray}}
\newcommand{\ena}{\end{eqnarray}}

\newcommand{\beano}{\begin{eqnarray*}}
\newcommand{\enano}{\end{eqnarray*}}

\newcommand{\bee}{\begin{enumerate}}
\newcommand{\ene}{\end{enumerate}}

\newcommand{\bei}{\begin{itemize}}
\newcommand{\eni}{\end{itemize}}

\newcommand{\betab}{\begin{tabular}}
\newcommand{\entab}{\end{tabular}}

\newcommand{\bd}{\begin{displaymath}}

\newcommand{\h}{{\mathfrak H}}

\newcommand{\hk}{{\mathfrak H}_{K}}

\newcommand{\bPhi}{\mbox{\boldmath $\Phi$}}
\newcommand{\bPsi}{\mbox{\boldmath $\Psi$}}

\newcommand{\hh}{\mbox{\boldmath $\mathfrak H$}}

\newcommand{\bcalN}{\mbox{\boldmath $\mathcal N$}}

\newcommand{\bK}{\mathbf K}
\newcommand{\bG}{\mathbf G}
\newcommand{\bh}{\mathbf h}

\newcommand{\bv}{\mathbf v}
\newcommand{\bE}{\mathbf E}

\newcommand{\bH}{\mathbf H}

\catcode `\@=11 \@addtoreset{equation}{section}

\catcode `\@=12

\begin{document}
\baselineskip 18pt plus 2pt minus 2pt

\begin{center}
{\LARGE \bf  Coherent  States on Hilbert Modules}\\[10mm]

{\large S. Twareque Ali\footnote{Supported in part by an NSERC grant}}\\

{\small Department of Mathematics and Statistics,  Concordia University}\\
{\small 1455 De Maisonneuve Blvd. West, Montr\'eal,}\\
{\small Qu\'ebec, Canada H3G 1M8}

\vspace{1cm}

{\large T. Bhattacharyya\footnote{Supported in part by DST
(Ramanna Fellowship) and UGC SAP Phase IV.}}\\
{\small Department of Mathematics, Indian Institute of Science,}\\
{\small Bangalore 560012, India}

\vspace{1cm}

{\large S. Shyam Roy\footnote{Supported in part by the
National Board for Higher Mathematics, India.}}\\
{\small Department of Mathematics and Statistics,}\\
{\small Indian
Institute of Science Education and Research, Kolkata}\\
{\small Mohanpur Campus, PO: BCKV Campus Main Office, Mohanpur
741252,}\\
{\small  Nadia, West Bengal, India}

\end{center}


\begin{abstract}
\baselineskip 18pt plus 2pt minus 2pt
  We generalize the concept of coherent states, traditionally defined as special
  families of vectors on Hilbert spaces, to Hilbert modules. We show that Hilbert modules over $C^*$-algebras are the natural
  settings for a generalization of coherent states defined on Hilbert spaces. We consider those Hilbert $C^*$-modules which have a natural left action from another
  $C^*$-algebra say, $\mathcal A$. The coherent states are well defined in this case and they behave well with respect to the left action by $\mathcal A$.
  Certain classical objects like the Cuntz algebra are related to specific examples of coherent states. Finally we show that coherent states on modules give
  rise to a completely positive kernel between two $C^*$-algebras, in complete
  analogy to the Hilbert space situation. Related to this there is a dilation result
  for positive operator valued measures, in the sense of Naimark. A number of examples are worked out to illustrate the theory.

\end{abstract}

\newpage


\section{Introduction}\label{sec:intro}
Coherent states (CS) are well-known objects in the physical literature. Mathematically,
they are an {\em overcomplete} set of vectors in a Hilbert space, labeled by points in some measure
space and satisfying  a certain {\em resolution of the identity} condition.
Although coherent states are defined and constructed in a variety of ways,
a general construction, and one which will be the prototypical
model for the generalization being proposed in this paper, may be described as
follows:
  Let $(X, \mu )$ be a finite measure space (often one requires that  $\mu (X) = 1$), with $X$
usually being a locally compact space, representing
the physical phase space of a classical mechanical system, the homogenous space
associated to some physical symmetry group, a coadjoint orbit of a locally compact
group, etc. Consider the Hilbert space $L^2 (X , d\mu )$ and let $\Phi_k ,\; k=0,1,2,3,
\ldots , N$ be an {\em orthonormal} set of vectors ($N$ is generally infinite, but it
could also be finite) in it,
 which satisfy the following condition,
 \be
   \mathcal N (x) := \sum_{k=0}^\infty \vert \Phi_k (x)\vert^2 < \infty\; , \quad
   x \in X\; .
\label{CS-cond}
\en
Coherent states are now defined, for each $x \in X$ to be the vectors,
\be
  \vert x \rangle = \sum_{k=0}^N \Phi_k\;\overline{\Phi_k (x)}\;\;
  \in L^2 (X , d\mu ),
\label{cs-def}
\en
Let $\h_K$ denote the subspace of $L^2 (X , d\mu )$ spanned by the $\Phi_k$. Then, the
following {\em resoluion of the identity} is easily established.
\be
   \int_X \vert x \rangle\langle x \vert\;d\mu (x) = I_{\h_K}\; .
\label{cs-resolid}
\en
It is also easily checked that $\h_K$ is a {\em reproducing kernel Hilbert space\/}, with
reproducing kernel $K (x, y ) = \langle x \mid y \rangle$ (see Section \ref{sec:examples}
below). A slight variant of this
construction is also used. Let $\mathfrak K$ be another Hilbert space of dimension
$N$ and $\{\psi_k\}_{k=0}^N$ an orthonormal basis of it. With the same vectors
$\Phi_k$ as before, we alternatively define coherent states as
\be
\vert x \rangle = \sum_{k=0}^N \psi_k\;\overline{\Phi_n (x)}\;\;
  \in \mathfrak K,
\label{cs-def2}
\en
which again satisfy a similar  resolution of the identity on $\mathfrak K$.

   This rather simple construction of an {\em overcomplete} family of vectors in a
Hilbert space, satisfying a resolution of the identity, turns out to be a powerful tool in many areas of physics and mathematics.
Detailed expositions of the theory of coherent states and their applications to
mathematics and physics may be found in \cite{aag-book,gaz,perel}.

   The purpose of the present paper is to suggest a possible construction of similar
overcomplete families of vectors in Hilbert $C^*$-modules (loosely speaking, Hilbert spaces over
$C^*$-algebras). We  shall call the resulting vectors {\em module valued coherent states
(MVCS)\/}. It is clear that since the field of complex numbers $\mathbb C$ is trivially a
$C^*$-algebra, coherent states on Hilbert spaces are special cases of MVCS. The richness of
the present generalization will be displayed with a number of examples. Some  definitions
and preliminary properties of Hilbert $C^*$-modules have been collected in
Appendix \ref{subsec:hilbmod}.

\section{Definition and construction of module valued coherent states}\label{sec:modCS}
  Let $\bE$ be an $\mathcal A - \mathcal B$ correspondence, where $\mathcal A$ and
$\mathcal B$ are unital $C^*$-algebras. This means that $\bE$ is a
Banach space which is  a Hilbert $C^*$-module over $\mathcal B$,
with a left action from $\mathcal A$, that is,  there is a
$*$-homomorphism from $\mathcal A$ into $\mathcal L (\bE)$. Note
that $\mathcal L (\bf E)$ denotes the bounded adjointable
operators on $\bE$ \cite[p. 8]{lan}. Note also that $\bE$ comes
equipped with a $\mathcal B$-valued inner product: $\langle
\cdot\mid \cdot\rangle_\bE:  \bE\times \bE\longrightarrow \mathcal
B$ given by $(f,g)\mapsto \langle f\mid g\rangle_\bE$ for $f,g \in
\bE$, which is antilinear in the first variable and linear in the
second. Furthermore, $\bE$ is complete in the norm $\Vert
f\Vert_\bE = [\Vert\langle f\mid f\rangle_\bE\Vert_{\mathcal
B}]^{\frac 12}$.
 Let $(X, \mu)$ be a finite measure space and consider the set of functions,
\be
  \mathbb F = \{ F : X \longmapsto \bE \mid  F \;\;\text{is a measurable function}\}\; . $$
  Then clearly, for any two $F, G$ in $\mathbb F$, $x \longmapsto
\langle F(x)\mid G(x)\rangle_\bE$ is a measurable function. Let
$$ \hh = \{ F \in \mathbb F \mid \mbox{ the function }
\langle F(x) \mid F(x) \rangle \mbox{ is Bochner integrable }\}\; .
\label{fcnsp}
\en

\belem $\hh$ is a complex vector space and an inner product module
over $\mathcal{B}$. \enlem

\prf That $\hh$ is a complex vector space follows from the fact
that a necessary and sufficient condition for $ \langle F(x) \mid
F(x) \rangle $ to be Bochner integrable is that $\int_X \Vert
\langle F(x) \mid F(x) \rangle_\bE
           \Vert_{\mathcal B}\; d\mu (x) < \infty$. Indeed, if $F,G \in \hh$,
then \beano
  \langle F(x) + G(x) \mid F(x) + G(x) \rangle_\bE   & =  &
      \langle F(x) \mid G(x)\rangle_\bE + \langle G(x) \mid F(x)\rangle_\bE \\
     &  + & \langle F(x) \mid F(x)\rangle_\bE +
     \langle G(x) \mid G(x)\rangle_\bE\; .\enano
But, we also know (see, for example \cite{lan}) that,
$$
 \Vert \langle G(x)\mid F(x)\rangle_\bE\Vert_\mathcal B \leq
 \Vert \langle G(x)\mid G(x)\rangle_\bE\Vert_\mathcal B^{\frac 12}\;
 \Vert \langle F(x)\mid F(x)\rangle_\bE\Vert_\mathcal B^{\frac 12}\; .$$
 By Schwarz inequality,
 \beano
 \int_X \Vert \langle G(x)\mid F(x)\rangle_\bE\Vert_\mathcal B\; d\mu (x)
  &\leq &  \int_X \Vert \langle G(x)\mid G(x)\rangle_\bE\Vert_\mathcal B^{\frac 12}\;
  \Vert \langle F(x)\mid F(x)\rangle_\bE\Vert_\mathcal B^{\frac 12}\;  d\mu (x)\\
  &\leq & \left( \int_X \Vert \langle G(x)\mid G(x)\rangle_\bE\Vert_\mathcal B\;
    d\mu(x)\right)\\
    &\times & \left( \int_X \Vert \langle F(x)\mid F(x)\rangle_\bE\Vert_\mathcal B\;
    d\mu(x)\right) < \infty\; . \enano
 Similarly,
 $$
 \int_X \Vert \langle F(x)\mid G(x)\rangle_\bE\Vert_\mathcal B\; d\mu (x) < \infty\; , $$
 so that
 $$
 \int_X \Vert \langle F(x) + G(x) \mid F(x) + G(x) \rangle_\bE\Vert_\mathcal B  \; d\mu (x)
  < \infty . $$
  In other words $F+G\in\hh.$ It is easy to see that $\hh$ is closed under multiplication by
complex scalars.

To make $\hh$  an inner product module over $\mathcal B$, we
define the right multiplication and the inner product respectively
by
$$ (F\cdot b)(x) = F(x)b\;\mbox{~for all~} b \in \mathcal B, \;\; \langle F\mid G\rangle_{\h} = \int_X \langle F(x)\mid
G(x)\rangle_\bE\; d\mu (x)\;  $$ on it. Then, for $b \in \mathcal
B$, \beano \langle F\mid G\cdot b \rangle_{\h}  & = & \int_X
\langle F(x) \mid G(x) b \rangle_\bE \;
         d\mu(x)\\
         & = & \int_X \langle F(x) \mid G(x)\rangle_\bE\; b \;  d\mu(x) \\
         & = & \int_X \langle F(x) \mid G(x)\rangle_\bE \;  d\mu(x)\;
         b =  \langle F \mid G\rangle_{\h}\; b\; .
\enano \qed

We have shown that  $\hh$ is an inner product $\mathcal B$-module, with respect to the
inner product and norm,
$$
\langle F\mid G\rangle_{\h} = \int_X \langle F(x)\mid G(x)\rangle_\mathbf E\;
d\mu (x)  \quad \text{and} \quad  \Vert F\Vert_\h = \Vert \langle F\mid \
F\rangle_\h\Vert_\mathcal B^{\frac 12}\; .  $$
Whenever there is an inner product $\mathcal B$-module, there are certain standard results which follow.
We collect these in the following lemma. The proofs can be found, for example, in \cite{lan}.
\begin{lem}
For $F, G\in \hh$ we have
\begin{enumerate}
\item $\langle G\mid F\rangle_\h\langle F\mid G\rangle_\h\leq\langle G\mid G\rangle_\h\Vert\langle F\mid F\rangle_\h\Vert_\mathcal B.$
\item $\Vert \langle F\mid G\rangle_{\h}\Vert_\mathcal B\leq \Vert F\Vert_\h\Vert G\Vert_\h.$
\item  $\Vert F+G\Vert_\h\leq \Vert F\Vert_\h+\Vert G\Vert_\h.$
\end{enumerate}
\end{lem}

\begin{lem}
$\hh$ is complete under the norm:
$$
    \Vert F\Vert_\h = \Vert \langle F\mid F\rangle_\bE\Vert_\mathcal B^{\frac 12}\; . $$
\end{lem}
\prf
Let $\{F_n\}$ be a Cauchy sequence in $\hh$. There is a subsequence $\{F_{n_i}\}, n_1<n_2<n_3\ldots$, such that
\begin{eqnarray}
\label{cauchy} \Vert F_{n_{i+1}}-F_{n_i}\Vert_\h< 2^{-i}\mbox{~for~}i\in \mb N.
\end{eqnarray}

Let us observe the following for $G\in \hh$
\begin{eqnarray}
\int_X\langle G(x)\mid G(x)\rangle_\bE d\mu(x) & \geq & \int_{\{x:\langle G(x)\mid G(x)\rangle_\bE\geq \epsilon\}}\langle G(x)\mid G(x)\rangle_\bE d\mu(x)\\
& \geq & \epsilon \mu(\{x:\langle G(x)\mid G(x)\rangle_\bE\geq \epsilon\})\\
\end{eqnarray}
Hence we have
\begin{eqnarray}
\label{cheby}\mu(\{x:\langle G(x)\mid G(x)\rangle_\bE\geq \epsilon\})\leq \epsilon^{-1}\int_X\langle G(x)\mid G(x)\rangle_\bE d\mu(x)
\end{eqnarray}

The Equation \eqref{cauchy} can be rewritten as
\begin{eqnarray}\Vert F_{n_{i+1}}-F_{n_i}\Vert^2_\h=\Vert\int_X\langle (F_{n_{i+1}}-F_{n_i})(x)\mid (F_{n_{i+1}}-F_{n_i})(x)\rangle_\bE d\mu(x)\Vert_\mathcal B< 2^{-2i}
\end{eqnarray}
Using the fact: if $a \in \mathcal B$ is positive and $\Vert a\Vert< \delta$, then $a<\delta$, we have
\begin{eqnarray}\label{norm}\int_X\langle (F_{n_{i+1}}-F_{n_i})(x)\mid (F_{n_{i+1}}-F_{n_i})(x)\rangle_\bE d\mu(x)< 2^{-2i}
\end{eqnarray}

Putting $A_i=\{x:\langle (F_{n_{i+1}}-F_{n_i})(x)\mid (F_{n_{i+1}}-F_{n_i})(x)\rangle_\bE\geq 2^{-i}\}$ and applying Equation \eqref{cheby} to $(F_{n_{i+1}}-F_{n_i})$ in place of $G$ we obtain from Equation \eqref{norm} that $$\mu(A_i)\leq 2^i\Vert F_{n_{i+1}}-F_{n_i}\Vert^2_\h<2^{-i}.$$ Hence we get $\sum_{i=1}^\infty \mu(A_i)<\infty .$ Now by Borel-Cantelli Lemma we have $\mu({\limsup}_n A_n)=0,$ where ${\limsup}_n A_n=\bigcap_{n=1}^\infty\bigcup_{k=n}^\infty A_k.$ But if $x\notin \limsup_n A_n $, then $\langle (F_{n_{i+1}}-F_{n_i})(x)\mid (F_{n_{i+1}}-F_{n_i})(x)\rangle_\bE<2^{-i}$ for large $i$. Equivalently, $\Vert\langle (F_{n_{i+1}}-F_{n_i})(x)\mid (F_{n_{i+1}}-F_{n_i})(x)\rangle_\bE\Vert_\mathcal B<2^{-i}$ for $x\notin \limsup_n A_n $. Hence $\{F_{n_i}(x)\}$ is a Cauchy sequence in $\bE$ for  $x\notin \limsup_n A_n$. Since $\bE$ is complete in the norm ${\Vert\langle .\mid.\rangle_\bE\Vert}^\frac{1}{2}_\mathcal B$ the sequence  $\{F_{n_i}\}$ converges pointwise almost everywhere to function $F$ (say). That is,
\begin{eqnarray} \label{pt}\lim_{i\rightarrow\infty}\Vert\langle (F_{n_{i}}-F)(x)\mid (F_{n_{i}}-F)(x)\rangle_\bE\Vert_\mathcal B=0\mbox{~for~} x\notin \limsup_n A_n.\end{eqnarray}
We define $F(x)=0$ for  $x\in \limsup_n A_n$. So $F$ is a function on $X$ such that Equation \eqref{pt} holds.
If $\mu$ is a finite measure then constants are integrable with respect to $\mu$. Hence by Dominated Convergence Theorem we conclude from Equation \eqref{pt} that
\begin{eqnarray}\label{dct} \lim_{i\rightarrow\infty}\int_X\Vert\langle (F_{n_{i}}-F)(x)\mid (F_{n_{i}}-F)(x)\rangle_\bE\Vert_\mathcal B d\mu(x)=0\end{eqnarray}
Since
\begin{eqnarray*}
\Vert F_{n_i}-F\Vert^2_\h &=& \Vert\int_X\langle (F_{n_{i}}-F)(x)\mid (F_{n_{i}}-F)(x)\rangle_\bE d\mu(x)\Vert_\mathcal B\\
&\leq & \int_X\Vert\langle (F_{n_{i}}-F)(x)\mid (F_{n_{i}}-F)(x)\rangle_\bE\Vert_\mathcal Bd\mu(x)\\
\end{eqnarray*}
It follows from Equation \eqref{dct} that
\begin{eqnarray*}\lim_{i\rightarrow\infty}\Vert F_{n_i}-F\Vert_\h=0.\end{eqnarray*}
Since $\{F_n\}$ is a Cauchy sequence we have
\begin{eqnarray*} \lim_{n\rightarrow\infty}\Vert F_{n}-F\Vert_\h=0.\end{eqnarray*}
Equation \eqref{dct}allows us to pick some $n_i$ such that
 $$\int_X\Vert\langle (F_{n_{i}}-F)(x)\mid (F_{n_{i}}-F)(x)\rangle_\bE\Vert_\mathcal Bd\mu(x)<\infty.$$
 Hence $F=(F-F_{n_i})+ F_{n_i} \in \h.$ Therefore $\hh$ is complete in the norm specified.
\qed
\medskip

  Note also that $\hh$ is an $\mathcal A - \mathcal B$ correspondence. This is so because  for
any $a \in \mathcal A$ and $F\in \hh$, we  may define $(a \cdot
F)(x) =  aF(x) $ where,  for $f \in \bE$, by $af$ we mean the left
action of $a$ on $f$, through its image in $\mathcal L (\bE)$ under
the postulated $*$-homomorphism. Moreover, \bea \langle a\cdot F\mid
G\rangle_\h & =& \int_X \langle a F(x)\mid G(x)\rangle_\bE\; d\mu
(x)
                            \nonumber\\
    & = &  \int_X \langle  F(x)\mid a^* G(x)\rangle_\bE\; d\mu (x)\; , \;\; \text{in view of
        the left action of $\mathcal A$ on $\bE$,} \nonumber\\
    & = & \langle F \mid a^*\cdot G\rangle_\h\; .
\label{adj}
\ena

At this point, let us introduce a notation which we shall use consistently in the sequel. For
$e \in \bE$, we define the map $ \langle e | : \bE \longrightarrow \mathcal B$, by
$$
  \langle e | (f) = \langle e\mid f\rangle_\bE \; , \quad f \in \bE\; . $$
  This is an {\em adjointable map\/}. We shall denote its adjoint by $| e \rangle $.
  Then  $ | e \rangle : \mathcal B \longrightarrow \bE$ has
the action
$$ | e \rangle (b) = eb \;,  \quad b \in \mathcal B\; , $$
so that for $e_1, e_2 \in \bE$,
\be
  | e_1 \rangle \langle e_2 | (f) = e_1 \langle e_2 \mid f\rangle_\bE\; .
\label{rank1op}
\en
Thus formally, one may use the standard ``bra-ket'' notation for Hilbert modules as one
does for Hilbert spaces.

\subsection{Non-normalized module valued CS}\label{subsec: non-norm-CS}
Proceeding now to construct coherent states,  we choose a set of  vectors $F_0, F_1, \ldots ,
F_n , \ldots\; $ (finite or infinite) in the function space $\hh$ (see (\ref{fcnsp})),  which are {\em pointwise defined}  (for all $x \in X$)
and  which satisfy the {\em orthogonality relations\/},
\be
   \int_X \mid F_k (x)\rangle \langle F_\ell (x)\mid \; d\mu (x)  = I_\bE\; \delta_{k \ell}\; .
    \label{functions}
\en
Next we take a second Hilbert module $\bG$, over another $C^*$-algebra
$\mathcal C$, which may or may not be the same algebra as $\mathcal B$. In $\bG$ we choose a
set of elements, $\phi_0, \phi_1, \ldots, \phi_n , \ldots$,  of the same cardinality as  of the
$F_k$, and which satisfy,
\be
\sum_k \mid \phi_k\rangle \langle \phi_k \mid = I_\bG.
\label{exframe}
\en
Note
that it follows from (\ref{exframe}), that any element $f \in \bG$ can be written as a
linear combination of the $\phi_k$, with  $\mathcal C$-valued coefficients:
$$
    f  =  \sum_k \mid \phi_k\rangle \langle \phi_k \mid (f) =  \sum_k \phi_k c_k\; ,
    \qquad c_k =  \langle \phi_k \mid f\rangle_\bG \in \mathcal C\; . $$

  Let $\bH = \bE \otimes \bG$ denote the {\em exterior tensor product} (see, for example, \cite{lan})
of the two Hilbert modules $\bE$ and $\bG$, which is then itself a Hilbert module over
$\mathcal B \otimes \mathcal C$. Here and elsewhere in the paper, we consider only
the projective tensor product of $C^*$-algebras which is also called special or minimal. In case one of our $C^*$-algebras is nuclear, all $C^*$-norms on the algebraic tensor product coincide and hence in that case there is a unique tensor product.
For each $x \in X$ and {\em co-isometry} $a \in \mathcal A$
(i.e., $aa^* = \text{id}_{\mathcal A}$), we define the vectors,
\be
  \mid x, a\rangle = \sum_k a F_k (x) \otimes \phi_k  \in \bH \; ,
\label{modCS}
\en
assuming of course that the sum converges. We call these vectors (non-normalized) {\em module valued
coherent states (MVCS)\/.}

\belem The MVCS in (\ref{modCS}) satisfy the resolution of the identity,
\be
\int_X \mid x, a\rangle \langle x, a\mid\; d\mu(x)  = I_\bH.
\label{modresolid}
\en
\enlem

\prf

It is enough to prove the identity on elements in $\bH$ of the type $h = e\otimes g$, with
$e \in \bE$ and $g\in \bG$. Since these elements form a total set in $\bH$, the lemma
will be proved by extending by continuity.  Indeed,
\beano
& & \left( \int_X \mid x, a\rangle \langle x, a\mid\; d\mu(x) \right)(e \otimes g) \\
  & = & \int_X \mid x, a \rangle \langle  \mid x, a \rangle \mid e\otimes g \rangle_\bH\; d\mu (x)\\
  & = & \int_X \sum_k (a F_k (x)\otimes \phi_k ) \cdot  \sum_\ell  \langle a F_\ell (x)\otimes \phi_\ell \mid
       e\otimes g\rangle_\bH \; d\mu (x)\\
  & = & \sum_{k, \ell}\int_X (a F_k (x)\otimes \phi_k )\; ( \langle a F_\ell (x) \mid e\rangle_\bE
 \otimes \langle \phi_\ell \mid g\rangle_\bG )\; d\mu(x)\\
  & = & \sum_{k, \ell}\int_X  a F_k (x)\langle a F_\ell(x)\mid e\rangle_\bE
        \otimes\phi_k \langle \phi_\ell \mid g\rangle_\bG\;d\mu(x)\\
  & = & \left(\sum_{k, \ell}\int_X | aF_k (x) \rangle \langle aF_\ell (x) | (e)\; d\mu (x)\right)
 \otimes\phi_k \langle \phi_\ell \mid g\rangle_\bG\; ,
   \quad   \text{by virtue of  (\ref{rank1op})}\\
  & = & \sum_{k, \ell}a\left(\int_X | F_k (x) \rangle \langle F_\ell (x) | \; d\mu (x)\right)a^*e
             ~\otimes \mid\phi_k\rangle\langle\phi_\ell\mid(g)
            \\
  & = & aa^* e\otimes \sum_{k} | \phi_k \rangle \langle  \phi_k | (g)\; , \quad \text{in view of
                   (\ref{functions})}\\
  & = & e\otimes g \; , \quad \text{by assumption (\ref{exframe})}\; .
\enano
This proves the resolution of the identity holds for vectors of the postulated type. The lemma is
proved, as stated earlier, by continuity. \qed

(Note that in the above expressions, we are using the same notation, $\otimes$, to denote
tensor products between different spaces. However, it is clear from the context which spaces are
meant in any given instance.)

\subsection{Normalized module valued CS}
We now proceed to show how the above construction may be modified to obtain MVCS which are
normalized.

In constructing the non-normalized MVCS two constraints,
(\ref{functions}) and (\ref{exframe})
 were imposed. We now impose two additional conditions, in order to obtain the
 {\em normalized} versions of these MVCS. We denote the normalized MVCS by
 $\mid\widehat{ x, a\rangle}$ and require that
 \be
    \widehat{\langle x, a }\mid \widehat{x, a  \rangle} = \text{id}_\mathcal B \otimes \text{id}_\mathcal C\; .
 \label{normal-cond}
 \en
 In order to achieve this we first require that
\be
  \langle \phi_k \mid \phi_\ell \rangle_\bG =\text{id}_\mathcal C\;  \delta_{k\ell}\;,
  \qquad \mbox{for all~} k, \ell\geq 0\; .
\label{orthonorm}
\en
Next we define
\be
 \mathcal N(x, a) = \langle x, a \mid x, a\rangle_\bH  = \sum_k \langle F_k (x) \mid a^*a
      F_k (x)\rangle_\bE \otimes \text{id}_\mathcal C\;,
\label{normal-cond2}
\en
the second equality following from (\ref {adj}) and    (\ref{orthonorm}).
We now require that for each $x \in X$,
\be
 \mathcal N(x, \text{id}_\mathcal A ) = \langle x, \text{id}_\mathcal A \mid x, \text{id}_\mathcal A \rangle_\bH  =
  \sum_k \langle F_k (x) \mid
      F_k (x)\rangle_\bE \otimes \text{id}_\mathcal C > 0\; ,
\label{normal-cond3}
\en
in the sense that we require the existence of $ \mathcal N(x, \text{id}_\mathcal C )$ as a
{\em positive invertible element} in the $C^*$-algebra $\mathcal B \otimes \mathcal C$.
Additionally, we require that $a$ be an invertible element in $\mathcal A$. In that case,
$$
\mathcal N(x, a ) > \inf_{\lambda \in \sigma (a^* a)} \lambda  \mathcal N(x, \text{id}_\mathcal C )
    > 0\; , $$
$\sigma (a^*a)$ denoting the spectrum of $a^*a$ in $\mathcal C$ and hence $\mathcal N(x, a )$
is invertible in $\mathcal C$.

  Finally, we define the {\em normalized MVCS} as,
\be
   \mid \widehat{x, a \rangle} = \mid x, a \rangle\; \mathcal N(x, a )^{-\frac 12}\; .
\label{normalizedMVCS}
\en
It is then straightforward to verify that these CS satisfy the normalization condition (\ref{normal-cond})
and the resolution of the identity,
\be
  \int_X  \mid \widehat{x, a \rangle}\mathcal N(x, a ) \widehat{\langle x,a}\mid\; d\mu(x)
   = I_\bH\; , \qquad \bH = \bE \otimes \bG\; ,
\label{rormMVCS-resolid}
\en
which should be compared to (\ref{modresolid}).

\section{Some examples}\label{sec:examples}
Let us look at a few examples which illustrate the above construction.

\medskip

\eg1. {\bf {\em Standard coherent states}}

First we show that the usual Hilbert space valued coherent states
are contained in our definition. As stated in the Introduction, these coherent states can
generically be obtained as follows. We start with the Hilbert space
$\h = L^2 (X, \mu)$ and assume that it contains a reproducing kernel
subspace, which we denote by $\hk$. This means that there exists an
integral kernel, $K: X \times X \longrightarrow \mathbb C$, which
satisfies $\overline{K(x,y)} = K(y,x)\; , \;\; K(x,x) >0, \;\;
\mbox{for all~} x,y\in X$, and for any $f \in \hk$,
$$
    f(x) = \int_X K(x,y)f(y)\; d\mu (y)\; , \qquad \mbox{for all~} x \in X\; . $$
    Moreover, if $\mathbb P_K$ denotes the projection operator from $L^2(X, \mu )$ to the subspace
$\hk$, then $K(x,y)$ is the integral kernel of this operator. If $\Phi_0 , \Phi_1 , \ldots ,
\Phi_n , \ldots$ is any orthonormal basis of $\hk$, then
\be
   K(x,y) = \sum_k  \Phi_k (x)\overline{\Phi_k (y)}\; .
\label{repker}
\en
Using this fact, one can define {\em non-normalized}  coherent states as
\be
    \mid x \rangle :  = K(\cdot , x ) =  \sum_k  \Phi_k\;\overline{\Phi_k (x)} \in \hk\; .
\label{hilbspCS}
\en
It is then easy to verify that
\be
 \langle x \mid y\rangle = K(x,y) \quad \text{and} \quad \int_X \mid x \rangle\langle x \mid \; d\mu (x)
      = I_{\mathfrak H_K}\; .
 \label{CS-props}
 \en
Generally, one can take any other Hilbert space $\mathfrak K$, the dimension of which has the same
cardinality as that of $\hk$ and define coherent states in $\mathfrak K$ as
$$
\mid x \rangle  =  \sum_k \psi_k\;\overline{\Phi_k (x)}  \; , $$
where $\psi_1 , \psi_2 , \ldots , \psi_n , \ldots$ is an orthonormal basis of $\mathfrak K$. These CS
satisfy both the conditions in (\ref{CS-props}), with $I_{\mathfrak H_K}$ replaced by $I_\mathfrak K$.
Furthermore,
since $K(x,x) = \sum_k \vert \Phi_k (x)\vert^2 := \mathcal N (x) >0$, normalized CS can be defined as:
$$
  \widehat{\mid x\rangle} =  \mathcal N(x)^{-\frac 12} \mid x \rangle\;, $$
which then satisfy, the conditions,
$$\Vert \widehat{\mid x \rangle}\Vert = 1 \quad \text{and} \quad \int_X \widehat{\mid x \rangle}
 \widehat{\langle x \mid} \; \mathcal N (x)\; d\mu (x)  = I_{\mathfrak K}\; .$$
 In order to arrive at these coherent states from our previous construction, we take $\mathcal A
 = \mathcal B = \mathcal C = \mathbb C$, also $\bE = \mathbb C, \; \bG = \mathfrak K$ (both
 considered as Hilbert modules over $\mathbb C$ and $F_k (x) = \overline{\Phi_k (x)}$.

 \medskip

 \eg2. {\bf {\em Vector coherent states}}

 Vector coherent states can generically be constructed as follows. Consider the Hilbert space
$\h = L^2_{\mathbb C^N} (X, \mu )$, of $\mathbb C^N$-valued functions on $X$, with scalar product
$$\langle \mathbf f\mid \mathbf g \rangle_\h = \int_X \mathbf f(x)^\dagger \mathbf g(x)\; d\mu(x)\; . $$
(In our notation, $\mathbf f(x)$ is the column vector with components $f_i (x)$ and
 $\mathbf f(x)^\dag$ is the row vector
$(\overline{f_1 (x)}, \overline{f_2 (x)}, \ldots , \overline{f_N (x)})$ .) Suppose there exists a
reproducing kernel subspace $\h_\bK \subset L^2_{\mathbb C^N} (X, \mu )$, with a (matrix valued)
kernel $\bK : X \times X \longrightarrow \mathcal M_N (\mathbb C)$ (set of all $N\times N$
complex matrices).  Denote by $\mathbb P_\bK$ the projection operator from $\h$ to $\h_\bK$ and
let $\bPhi_0 , \bPhi_1 , \ldots , \bPhi_n , \ldots ,$ be any orthonormal basis of $\h_\bK$. Then,
\bea
  \bK (x, y ) & = & \sum_k \bPhi_k (x) \bPhi_k (y)^\dag\; , \qquad \bK (x, y )^* = \bK (y, x)\; , \; \;
  \; \forall x,y \in X\\
  (\mathbb P_\bK \mathbf f )(x) & = &\int_X \bK (x, y) \mathbf f (y)\; d\mu (x)\; ,
     \quad \mathbf f \in \h\; .
\label{matrix-ker}
\ena
Furtheremore, for each $x \in X , \; \bcalN (x) := \bK (x,x)  =  \sum_k \bPhi_k (x) \bPhi_k (x)^\dag$ is
a positive, invertible matrix  and
$$
   \int_X  \bK (x, z) \bK (z, y) \; d\mu (z) = \bK (x, y) \; . $$
Let $\Phi_k^1 (x), \Phi_k^2 (x), \ldots ,  \Phi_k^N (x)$ denote the components of the $N$-vector
$\bPhi_k (x)$ and let $\{\chi^i\}_{i=1}^N$ be an orthonormal basis of $\mathbb C^N$. Vector
coherent states (VCS) are now defined to be the elements in $\h_\bK$:
\be
  \mid x, i\rangle : = \bK (\cdot , x )\chi^i =  \sum_k \bPhi_k  \overline{\Phi_k ^i (x)}\; , \quad x \in X,
     \;\; i = 1, 2, \ldots , N\; .
\label{VCS}
\en
These satisfy the conditions,
$$
   \langle x , i \mid y, j \rangle = \bK (x, y)_{ij}\; , \qquad
      \sum_{i=1}^N \int_X \mid x, i\rangle \langle x, i\mid \; d\mu (x) = I_{\h_{\bK}}\; . $$
This time,  ``normalized'' VCS are  defined as,
$$
 \mid \widehat{x, i \rangle} = [ \text{Tr}( \bcalN (x))]^{-\frac 12}\mid x, i \rangle \; \quad
 \text{so that,} \quad \sum_{i=1}^N \Vert \mid x, i \rangle\Vert^2 = 1\; , $$
and
$$
 \sum_{i=1}^N \int_X  \mid \widehat{x, i \rangle}
  \widehat{\langle x, i }\mid \;   \text{Tr}( \bcalN (x)) \; d\mu (x) = I_{\h_{\bK}}\; . $$

    We now show how these VCS can be associated, and in fact obtained, from a family of MVCS.
This will be achieved by extending the space $X$ over which the coherent states are
defined. Take the group $SU(N)$ of $N\times N$ unitary matrices with determinant one. Let
$d\Omega$  denote its Haar measure (normalized to one).  It is known from the general theory
of compact groups that for any  normalized vector  $\bv \in  \mathbb C^N$, one has the
relation,
\be
   \int_{SU(N)} u\bv \bv^\dag u^* \; d\Omega (u) = \frac 1N \mathbb I_N\; .
\label{sun-reln}
\en
Consider now the domain $X \times SU(N)$ and the orthonormal basis $\{\bPhi_i\}$ of the
reproducing kernel Hilbert space $\h_\bK$, considered above. Let us now define the
matrix valued functions $F_k : X\times SU(N) \longrightarrow \mathcal M_N (\mathbb C)$,
\bea
   F_k (x, u ) & = &  N^{\frac 12}\;u\;  \text{diag}[\;\overline{\Phi^1_k(x)} , \;
   \overline{\Phi^2_k(x)} , \;\ldots , \;\overline{\Phi^N_k (x)}\;]\;  u^* \nonumber\\
    & = &
     N^{\frac 12} \sum_{i=1}^N u \; \overline{\Phi^i_k(x)}\;\mathbb P_i \; u^* \; , \qquad
     (x, u) \in X\times SU(N)\; ,
\label{matrix-fcns}
\ena
where the $\mathbb P_i$ are the one-dimensional projection operators, $\chi^i \chi^{i\dag}$,
built out of the vectors $\chi^i$ in the chosen orthonormal basis of $\mathbb C^N$.
It is then not hard to see, using the orthonormality of the vectors $\{\bPhi_i\}$ and the relation
(\ref{sun-reln}), that
\be
   \int_{X\times SU(N)} F_k (x) F_\ell (x)^*\; d\mu (x )\; d\Omega (u) = \mathbb I_N\; \delta_{k\ell}\; ,
     \qquad k, \ell = 1,2, \ldots , N\; .
\label{matrix-fcns2}
\en
Referring to the general construction of MVCS in Section \ref{subsec: non-norm-CS}, we take
$\mathcal A = \mathcal B = \mathcal M_N (\mathbb C)$ and $\bE = \mathcal M_N (\mathbb C)$,
considered as a Hilbert module over itself. We take $\bG$ to be the Hilbert space $\h_\bK$,
considered as a Hilbert module over $\mathbb C$.
The MVCS are then defined as:
\be
   \mid x, u , V\rangle = \sum_k VF_k (x, u) \otimes\bPhi_k \in \bH =
   \mathcal M_N (\mathbb C)\otimes \h_\bK\; ,  \quad \mbox{for all~} (x, u)\in
                                X\times SU(N)\; ,
\label{MVCS-VCS}
\en
where $V$ is a unitary element in $SU(N)$. These MVCS satisfy the resolution of the identity,
$$
 \int_{X\times SU(N)}\mid x, u , V\rangle \langle x, u , V\mid\; d\mu (x)\; d\Omega(u) =
          I_\bH\; .$$

  In order to recover the VCS (\ref{VCS}) from here, we use the projection operators,
$\mathbb P_i (u) = u\mathbb P_i u^*\; , u \in SU(N)$ and simply take the partial trace
in $\mathcal B = \mathcal M_N (\mathbb C)$,
\be
  \mid x , i \rangle = \text{Tr}_{\mathcal B}[\;\mathbb P_i (u)\mid x, u,
  \mathbb I_N\rangle \;]\; .
\label{MVCStoVCS}
\en

   A related example is that of the analytic VCS, built in \cite{alieng}, using powers of
matrices from $\mathcal M_N (\mathbb C)$. These VCS may be defined as:
\be
  \mid \mathfrak Z , i \rangle =  \sum_k \frac {{\mathfrak Z}^k}{\sqrt{c_k}}\chi^i \otimes \bPhi_k\; \qquad
              \mathfrak Z \in  \mathcal M_N (\mathbb C)\; ,
\label{anal-vcs}
\en
where the $c_k$ are the numbers (see, e.g. \cite{alieng,gin}],
$$
  c_k  =  \frac 1{(k+1)(k+2)}\left[\prod_{j = 1}^{k+1}(N +j) - \prod_{j = 1}^{k+1}(N -j)\right],
   \qquad   k =0, 1, 2, \ldots ,
$$
Let $z_{ij}, \; i,j = 1,2, \ldots , N$ be the matrix elements of $\mathfrak Z$. Then, writing
$$F_k (\mathfrak Z ) = \frac {\mathfrak Z^k}{\sqrt{c_k}}\quad \text{and} \quad
   z_{ij} = x_{ij} + i y_{ij}\;,$$
it can be shown that,
$$
\int_{\mathcal M_N (\mathbb C)} F_k (\mathfrak Z ) F_\ell(\mathfrak Z)^*\; d\mu (\mathfrak Z, \mathfrak Z^*)
   = \delta_{k\ell}\mathbb I_N\; , \qquad
 d\mu (\mathfrak Z, \mathfrak Z^*)  = \frac {e^{-\text{Tr}[\mathfrak Z^* \mathfrak Z]}}{{(2\pi)}^N}\;
 \prod_{i,j = 1}^N dx_{ij}\; dy_{ij}\;. $$
Using this fact, one may prove the resolution of the identity,
$$
\sum_{i=1}^N\int_{\mathcal M_N (\mathbb C)} \mid \mathfrak Z , i\rangle
   \langle \mathfrak Z, i\mid\; d\mu (\mathfrak Z , \mathfrak Z^* ) = \mathbb I_N \otimes
    I_{\h_\bK}\; .
$$

   To construct the related MVCS, we consider    $\mathcal M_N (\mathbb C)$ as a module over
itself and identify it with
$\bE$. The module $\hh$, containing the functions $F_k$, then consists of functions from
$\mathcal M_N (\mathbb C)$ to itself.  Considering  $\h_\bK$ as a
module over  $\mathbb C$, we may define MVCS in $\bH =
\mathcal M_N (\mathbb C)\otimes \h_\bK$ as
\be
\mid \mathfrak Z , a \rangle = \sum_k a F_k (\mathfrak Z ) \otimes \bPhi_k =
  \sum_k a\frac {{\mathfrak Z}^k}{\sqrt{c_k}} \otimes \bPhi_k\; ,
\label{anal-mvcs}
\en
where $a$ is a unitary element in $\mathcal M_N (\mathbb C)$. These then satisfy the resolution
of the identity,
\be
  \int_{\mathcal M_N (\mathbb C)} \mid \mathfrak Z , a\rangle
   \langle \mathfrak Z, a\mid\; d\mu (\mathfrak Z , \mathfrak Z^* ) = I_\bH\; .
\label{anal-mvcs-resolid}
\en

 In the particular case when $N = 2$ the set $\mathcal M_N (\mathbb C)$, of all
complex $2\times 2$ matrices,  can be identified with the space of {\em complex
quaternions\/}. The resulting MVCS may then be called {\em complex quaternionic MVCS\/}.

\medskip
\eg3. {\em{\bf A real quaternionic variant}}

   Vector coherent states of the type (\ref{anal-vcs}), when $\mathfrak Z$ is replaced by a
{\em real} quaternionic variable $\mathfrak q$, have been constructed in \cite{AEG-VCS} and \cite{thirali}, while
coherent states in quaternionic Hilbert spaces have been studied in \cite{admil}.
These latter coherent states, which are a natural generalization of the canonical coherent states
to  quaternionic quantum mechanics \cite{adler-quat-qm}, have also been shown to have
interesting physical applications. We now construct an analogous family of quaternionic coherent
states on a quaternionic Hilbert space. Recall that  a quaternionic Hilbert space is a linear
vector space over the field of (real) quaternions, $\mathbb H$, with  the inner
product taking values in $\mathbb H$. While  $\mathbb H$ contains the complexes, it is
not a $C^*$-algebra. So strictly speaking, a quaternionic Hilbert space is not a Hilbert
$C^*$-module. However, the quaternionic CS we shall now construct are very similar to
the MVCS  (\ref{anal-mvcs}).

   Let us start with the quaternionic vector coherent states
introduced in \cite{thirali}. These are vector coherent states defined on a standard
Hilbert space $\h$ (over the complexes). We take for $\mathfrak q \in \mathbb H$  its
representation by $2\times 2$ complex marices:
\be
\mathfrak q = u(\theta , \phi) \begin{pmatrix} z & 0 \\ 0 & \overline{z}\end{pmatrix}
u(\theta, \phi)^*\; , \qquad  u (\theta , \phi ) =  \begin{pmatrix}
ie^{i\frac {\phi}2}\cos\frac {\theta}2 &  - e^{i\frac {\phi}2}\sin\frac {\theta}2\\
e^{-i\frac {\phi}2}\sin\frac {\theta}2 &  - ie^{-i\frac {\phi}2}\cos\frac {\theta}2\; ,
\end{pmatrix}
\en
\label{quaternion1}
where $z \in \mathbb C , \; 0 \leq \theta \leq \pi , \; 0 < \phi \leq 2\pi $.  Writing $z = re^{i\xi}$,
we also have,
\be
   \mathfrak q = r [\mathbb I_2 \cos \xi + i \sigma (\widehat{n})\sin \xi ]  =
       r e^{i\xi \sigma (\widehat{n})},
\label{quaternion2}
\en
with
$$
 \mathbb I_2 = \begin{pmatrix} 1 & 0 \\ 0 & 1
\end{pmatrix}, \qquad   \sigma (\widehat{n}) = \begin{pmatrix} \cos\theta & e^{i\phi}\sin\theta \\
e^{-i\phi}\sin\theta   & -\cos\theta \end{pmatrix},
\qquad [\sigma(\widehat{n})]^2 = \mathbb I_2 $$

   Let $\{\bPsi_n\}_{n=0}^\infty$ be an orthonormal basis of  $\h$ and
$\chi^i, \; i = 1,2$, an orthonormal basis of $\mathbb C^2$. Normalized {\em quaternionic vector coherent
states} are then defined \cite{AEG-VCS,thirali} as
\be
\mid \mathfrak q , j \rangle = \frac {e^{-\frac {r^2}2}}{\sqrt{2}} \sum_{n=0}^\infty
  \frac {\mathfrak q^n}{\sqrt{n!}}\chi^i\otimes \bPsi_n \; \in \mathbb C^2 \otimes \h\;,
  \qquad \sum_{j=1}^2\Vert \mid \mathfrak q , j \rangle\Vert^2 = 1 \; .
\label{quaternion-vcs}
\en
These vectors satisfy the resolution of the identity,
\be
  \sum_{j=1}^2\int_{\mathbb H} \mid \mathfrak q , j \rangle \langle \mathfrak q , j \mid\;
     d\mu (\mathfrak q , \mathfrak q^\dag ) = \mathbb I_2 \otimes I_\h\; , \qquad
     d\mu (\mathfrak q , \mathfrak q^\dag ) = \frac 1{8\pi^2} \;rdr\;d\xi \;
     \sin\theta d\theta\;d\phi\; .
\label{quater-resolid1}
\en

  Suppose now that $\hh_{\rm quat}$ is a Hilbert space over the quaternions. (Multiplication
by elements of $\mathbb H$ from the right is assumed, i.e., if $\bPhi \in \hh_{\rm quat}$  and
$\mathfrak q \in \mathbb H$, then $\bPhi \mathfrak q \in \h_{\rm quat}$). The obvious
generalization of the VCS (\ref{quaternion-vcs}) to {\em quaternionic coherent states} over
$\hh_{\rm quat}$ are easily written down by taking an orthonormal basis
$\{\bPsi_n^{\rm quat}\}_{n=0}^\infty$ in $\hh_{\rm quat}$ and defining the vectors
\be
  \mid \mathfrak q \rangle = e^{-\frac {r^2}2}\sum_{n=0}^\infty \bPsi_n^{\rm quat}
  \frac {\mathfrak q^n}{\sqrt{n!}}\; \in  \hh_{\rm quat}, \qquad \mathfrak q \in
  \mathbb H, \qquad \langle \mathfrak q \mid \mathfrak q \rangle_{\h_{\rm quat}}
   = \mathbb I_2\; .
\label{quater-cs}
\en
They satisfy the resolution of the identity,
\be
  \int_{\mathbb H} \mid \mathfrak q \rangle \langle \mathfrak q \mid\;
     d\nu (\mathfrak q , \mathfrak q^\dag ) =  I_{\h_{\rm quat}}\; , \qquad
     d\nu (\mathfrak q , \mathfrak q^\dag ) = \frac 1{4\pi^2} \;rdr\;d\xi \;
     \sin\theta d\theta\;d\phi\; .
\label{quater-resolid2}
\en

These coherent states were obtained in \cite{admil}, where a group theoretical argument
was used to construct them. Recently they have also been obtained in
\cite{thihonkry}. Here we stress their similarity with our general construction
over $C^*$-modules.

\medskip
\eg4. {\em{\bf Infinite component VCS}}

As a similar example to the above, but this time involving VCS with an infinite number of
components, we consider the VCS
\be
   \mid z , \overline{z}'; \ell \rangle = e^{-\frac 12(\vert z' \vert^2+ \vert z \vert^2)}\; \overline{z'}^\ell
      \sum_{n=0}^\infty \frac {z^n}{\sqrt{n! \ell !}}\mid \bPsi_{n}\rangle \; ,
      \qquad \ell = 0, 1,1 2, \ldots , \infty\; ,\;\;\; (z , \overline{z}') \in \mathbb C
        \times \mathbb C\; .
\label{landau-cs1}
\en
where the $\bPsi_{n}$ form an orthonormal basis in some Hilbert space $\h$. These
VCS are similar to those appearing in the problem of an electron moving in a constant magnetic field and its
associated {\em Landau levels\/} \cite{alibag}. They satisfy the normalization condition,
$$
  \sum_{\ell=0}^\infty \langle z , \overline{z}'; \ell \mid z , \overline{z}'; \ell \rangle = 1 \; ,
$$
and the resolution of the identity,
\be
   \sum_{\ell =0}^\infty\int_{\mathbb C} \mid z , \overline{z}'; \ell \rangle \langle z , \overline{z}'; \ell \mid
   \; \frac{dx\;dy}\pi = I_\h \;, \qquad  z = x + iy\; .
\label{vcs-resolid2}
\en

  In order to construct a family of MVCS corresponding to this set of VCS, we start with a
locally compact, unimodular group $G$  (such as, e.g., $SU(1,1)$), which has a representation,
in the discrete series, in an {\em infinite dimensional} Hilbert space $\mathfrak K$. Let
$G\ni g \mapsto U(g)$ be such a unitary irreducible representation and let
$d\mu_G$ denote the Haar measure of $G$. It is then well-known (see, e.g., \cite{aag-book})
that if  $\phi$ is any unit vector in $\mathfrak K$, then
\be
  \frac 1d \int_G U(g) \mid \phi\rangle \langle\phi\mid  U(g)^* \; d\mu_G (g) = I_\mathfrak K\; ,
\label{discr-ser-id}
\en
where $d > 0$ is a constant, called the {\em formal dimension} of the representation $U$. Let
$\{\phi_i\}_{i=1}^\infty$ be an orthonormal basis of $\mathfrak K$ and $\mathbb P_i =
\mid \phi_i \rangle\langle \phi_i\mid$ the corresponding one-dimensional projection
operators. We define the functions, $F_k : \mathbb C \times \mathbb C\times G  \longrightarrow
\mathcal L (\mathfrak K )$ :
\be
  F_k (z, \overline{z}' ,  g ) = \frac 1{d^{\frac 12}}\;e^{-(\vert z'\vert^2+\vert z'\vert^2)}
        \frac {z^k}{\sqrt{k!}}
        \sum_{n = 1}^\infty \frac {\overline{z}^{\prime n}}{\sqrt{n!}}\;\mathbb P_n (g)\; ,
        \qquad \mathbb P_n (g) = U(g)\mathbb P_n U(g)^*\; .
\label{inf-fcns}
\en
It is then easy to see that,
$$
\int_{\mathbb C \times G} F_k (z, \overline{z}' ,  g )F_\ell (z, \overline{z}' ,  g )^*\;
    \frac{dx\;dy}\pi\;d\mu_G (g) =  \delta_{k\ell}\; I_\mathfrak K\; , \quad
     z = x +iy\; . $$

      Thus, considering $ \mathcal L (\mathfrak K )$ as a $C^*$-algebra and as a
Hilbert module over  itself, we again define the MVCS on $\bH = \mathcal L (\mathfrak K )
\otimes \h$,
$$
 \mid z, \overline{z}', g ;\; a \rangle  = \sum_{k=1}^\infty aF_k (z, \overline{z}' ,  g ) \otimes \bPsi_k\
 = \frac 1{d^{\frac 12}}\;e^{-(\vert z'\vert^2+\vert z'\vert^2)}\sum_{k, n}
   \frac {\overline{z}^{\prime n}\;{z}^k}{\sqrt{n!\; k!}}\;a \mathbb P_n (g)
 \otimes \bPsi_k
  ; ,
$$
where, once more,  $a$ is a unitary element in $ \mathcal L (\mathfrak K )$ and $\{\bPsi_k\}_{k=1}^\infty$
an orthonormal basisof $\h$.  These MVCS
clearly have all the required properties, e.g., the resolution of the identity,
$$
   \int_{\mathbb C \times G} \mid z, \overline{z}', g;\;
   a \rangle \langle  z, \overline{z}', g ;\; a \mid\;
      \frac{dx\; dy}\pi \; d\mu_G (g) = I_\bH\;  , $$
and the VCS can be obtained from them by taking the partial trace in $\mathcal L (\mathfrak K )$ :
$$
   \mid z, \overline{z}', \ell  \rangle = \text{Tr}_{\mathcal L (\mathfrak K )}\;[\;\mathbb P_\ell (g)
      \mid z, \overline{z}', g;\; I_{\mathfrak K}  \rangle\; ]\; . $$

In the next example we  construct a different variant of these MVCS, using Cuntz
algebras.

\medskip

\eg5. {\bf Coherent states from representations of  Cuntz algebras}

Let $S_1, S_2, \ldots$ be isometries on a complex separable Hilbert
space $\mathcal{K}$ (necessarily infinite dimensional) such that
$$\sum_{j=1}^\infty S_j S_j^* = I_{\mathcal{K}}$$
where the sum converges in the strong operator topology of $\mathcal{B}(\mathcal{K})$.
Multiplying both sides by $S_i^*$, we get
$$ S_i^* + S_i^* \sum_{j \neq i} S_j S_j^* = S_i^*$$
so that
$$ S_i^* \sum_{j \neq i} S_j S_j^* = 0 $$
But $ \sum_{j \neq i} S_j S_j^* $ is the projection onto the closure of the span of the
ranges of $S_j$ for $j \neq i$. So the range of $S_i$ is orthogonal to the range of
$S_j$ for all $j \neq i$. This is a representation of the Cuntz algebra
$\mathcal{O}_\infty$ with infinitely many generators. Take $\bG =
\mathcal{C}$ to  be the $C^*$-algebra generated by the
isometries $S_1, S_2, \ldots $. Choose $\phi_i = S_i$. Then
$$ \langle \phi_i , \phi_j \rangle = S_i^* S_j = \delta_{ij} I_{\mathcal{K}}
\mbox{ and } \sum_{k=1}^\infty | \phi_k \rangle \langle \phi_k | = I.$$
Our coherent states are
$$ | x, a \rangle = ( \sum_{k=1}^\infty  a \cdot F_k(x) \otimes S_k)
(\bcalN (x)^{-1/2} \otimes I).$$

We now construct an explicit example of a Cuntz algebra.
Let $\omega: \mathbb N^{>0} \longrightarrow \mathbb N^{>0} \times \mathbb N^{>0}$ be
a bijection ($\mathbb N^{>0}$
denoting the set of non-zero, positive integers). Consider
a Hilbert space $\h$ and let $\{\phi_n\}_{n\in \mathbb N^{>0}}$ be an orthonormal basis
of it. Writing
$\omega(n) = (k,\ell)$ we define a re-transcription of this basis in the manner
\be
    \psi_{k \ell} := \phi_n = \psi_{\omega(n)}\; , \qquad k, n, \ell \in \mathbb N^{>0}\; .
\label{bijection}
\en
Note that the two sets of vectors are exactly the same and satisfy,
$\langle \phi_m\mid \phi_n \rangle_\h
= \delta_{mn}$ and $\langle \psi_{mn}\mid \psi_{k \ell} \rangle_\h = \delta_{mk}\;\delta_{n\ell}$,
respectively.
Define the family of isometries $S_k , \; k \in \mathbb N^{>0}$ on $\h$, in the manner
\be
    S_k \phi_n = \psi_{k n}\; , \qquad n \in \mathbb N^{>0}\; .
\label{isometries1}
\en
Note that this defines an isometry and not a unitary map. Indeed, one has,
\be
S_k^* S_\ell = \delta_{k\ell}\;I_\h  \quad \text{and} \sum_{k\in \mathbb N^{>0}} S_k S_k^* =
    \sum_{k\in \mathbb N^{>0}}\mathbb P_k = I_\h\; ,
\label{isometries2}
\en
$\mathbb P_k$ being the projection operator onto the subspace $\h_k$ of $\h$ spanned by the vectors
$\psi_{k\ell}, \; \ell \in \mathbb N^{>0}$. Moreover, $S_k S_\ell^*$ is a partial isometry from $\h_\ell$
to $\h_k$.

   The $C^*$-algebra $\mathcal O_\infty$, generated by these isometries, is then a
Cuntz algebra. An explicit example of such a bijection $\omega$ is  given in
Appendix \ref{bij}.

   The above construction has an immediate application to a physical situation. We consider
the non-normalized version (with $a$ set to the unit element of $\mathcal A$),
$$ | x \rangle =  \sum_{k=1}^\infty F_k(x) \otimes S_k .$$
Let $X = \mathbb C$ and $\bE = L^2 (\mathbb C,\; \frac {e^{-\vert z\vert^2}}{2\pi}\; dx\; dy),
\;\; z = \frac 1{\sqrt{2}}(x + iy)$. We take  $F_k : \mathbb C \longrightarrow \mathbb C$ to be the functions,
$$
  F_k (z ) = \frac {z^{k-1}}{\sqrt{(k-1)!}}\; , \qquad k =1,2,3, \ldots \; . $$
Let $\psi_{k\ell}$ be the complex Hermite polynomials,
\be
 \psi_{k \ell} (\overline{z} , z)  = \frac {(-1)^{n+k-2}}{\sqrt{(\ell-1)! (k-1)!}}
 \;e^{\vert z\vert^2}\partial^{\ell-1}_{\overline z}\partial^{k-1}_z
   \; e^{-\vert z\vert^2}\; , \qquad  k, \ell = 1,2,3, \ldots \; .
\label{comp-herm-poly}
\en
These form an orthonormal basis of $L^2 (\mathbb C,\; \frac {e^{-\vert z\vert^2}}{2\pi}\; dx\; dy)$. The coherent states now become
\be
 \vert z \rangle =  \sum_{k=1}^\infty  \frac {z^{k-1}}{\sqrt{(k-1)!}} S_k  \; .
\label{land-cs2}
\en
Let  $\phi_n$ be as in (\ref{bijection}) and consider the vectors
$$
   \xi_{\overline{z}^\prime,\; n} = \frac {\overline{z}^{\prime n -1}}{\sqrt{(n -1)!}}
    \phi_n\; . $$
Then the vectors (in $L^2 (\mathbb C,\; \frac {e^{-\vert z\vert^2}}{2\pi}\; dx\; dy)$),
\be
\vert z, \overline{z}^\prime, n\rangle =
   \sum_{k=1}^\infty  \frac {z^{k-1}}{\sqrt{(k-1)!}} S_k \xi_{\overline{z}^\prime,\; n}
= \overline{z}^{\prime n -1}\sum_{k=1}^\infty \frac {z^{k-1}}{\sqrt{(k-1)! \;(n-1) !}}\;
   \psi_{k n} \;,
\label{vcs-h1}
\end{equation}
($\ell =  1,2,3, \ldots , \infty,$) are just the non-normalized versions of the
infinite component vector CS associated to the Landau levels, found in
\cite{alibag}.

\section{Reproducing kernel, carrier space and a minimal dilation}\label{sec:repker-carriersp}

We start with a brief description of a {\em completely positive kernel\/}.
Given two $C^*$-algebras $\mathcal A$ and $\mathcal B$, an
        $(\mathcal A,\mathcal B)${\em -reproducing kernel correspondence} on a set
        $X $, is an $(\mathcal A,\mathcal B)$-correspondence $\bE$ whose
        elements are $\mathcal B$-valued functions $f \colon (x,a) \mapsto
        f(x,a) \in
        \mathcal B$  on $X \times \mathcal A$. It is a vector space with respect to
        the usual pointwise vector space operations. Moreover,  there is a {\em kernel
        element} $k_{x} \in E$
        such that
        \begin{equation}  \label{repro-prop}
      f(x,a) = \langle  k_{x} , \, a \cdot f \rangle_{E}
        \end{equation}
        for every $x \in         X$.
        When this is the case we say that the function $K \colon X
        \times X \to \mathcal L(\mathcal A, \mathcal B)$ given by
        \begin{equation} \label{RKC-cpkernel}
     K(x, y)[a] = k_{x}(y,a)
        \end{equation}
        is the {\em reproducing kernel} for the reproducing kernel
        correspondence $E$.

{}From the inner product characterization in (\ref{repro-prop}) of the
point evaluation for elements in an $(\mathcal A,\mathcal B)$-reproducing kernel
correspondence $E$ on $X$ one easily deduces that the left $\mathcal A$-action
and the right $\mathcal B$-action are given by
        \begin{equation} \label{RKC-Aaction}
       (a \cdot f)(x', a') = f(x', a' a) \text{ and }  (f \cdot
       b)(x',a')=f(x',a')b.
       \end{equation}

       The mapping from $\mathcal A$ to $\mathcal B$ given by $a \mapsto
        f(x,a)$ is $\mathcal A$-linear for each fixed $f \in E$ and $x \in
        X$.  Since all our algebras are unital, this follows from
        the general identity $f(x,a) = (a \cdot f)(x,1_\mathcal A)$ (a
        consequence of \eqref{repro-prop}, \eqref{RKC-Aaction} and the linearity of
        the point-evaluation map $f \mapsto f(x,1_\mathcal A)$).  Note
        also that we recover the element $k_{x}$ from $K$ by
using formula
        \eqref{RKC-cpkernel} to define $k_{x}$ as a function of
        $(y,a)$ for
        each $x \in X$.

 Given a reproducing kernel $(\mathcal A, \mathcal B)$-correspondence, one can
 show that the associated
        reproducing kernel function $K \colon X \times X \to
        \mathcal L(\mathcal A, \mathcal B)$ defined by \eqref{RKC-cpkernel} is a
        completely
        positive kernel in the sense of \cite{BBLS}, i.e., the function
        $$
        ((x,a),(x',a'))\to K(x, x')[a^{*}a' ]
        $$
        is a positive $C^*$-algebra valued kernel. This means that
$$ \sum_{i,j=1}^{N} b_i^* K(x_{i}, x_{j})[a_{i}^{*} a_{j}]b_j$$
        is a positive element of $\mathcal B$
        for each choice of finitely many $(x_{1},a_{1}), \dots,
        (x_{N},a_{N})$ in $X \times \mathcal A$ and $b_1,\ldots,b_N$ in $\mathcal B$.
        The following theorem, which again can be found in \cite{BBFtH} gives a
        complete clarity.

        \begin{theorem}
        Given a function $K \colon X \times X \to
        \mathcal L(\mathcal A,\mathcal B)$, the following are equivalent:
        \begin{enumerate}
        \item $K$ is a completely positive kernel in the sense that the
        function from $(X \times \mathcal A) \times (X \times \mathcal A) \to
        \mathcal L (\mathcal A, \mathcal B)$ given by
        $$
     ((x,a),(x',a')) \mapsto K(x', x)[a^{*}a']
        $$
        is a positive kernel in the sense that
        $$
        \sum_{i,j=1}^{N}b_i^*K(x_i, x_j)[{a_i}^*a_j]b_j \ge 0 \text{ in } \mathcal B $$
 for all  $(x_{1}, a_{1}), \dots, (x_{N}, a_{N}) \in
        X \times \mathcal A\text{ and }b_1,\ldots b_N\in\mathcal B.
        $

        \item $K$ has a Kolmogorov decomposition in the sense of
        \cite{BBLS}, i.e., there exists an
        $(\mathcal A,\mathcal B)$-correspondence $E$ and a mapping $x \mapsto
k_{x} $
        from $X$ into $E$ such
        that
        \begin{equation*} 
     K(x,y)[a] =\langle  k_{x}, \, a \cdot k_{y} \rangle_{E}
     \text{ for all } a\in \mathcal A.
     \end{equation*}

        \item $K$ is the reproducing kernel for an $(\mathcal A,\mathcal B)$-reproducing
        kernel correspondence $E = E(K)$, i.e., there is an $(\mathcal A,
        \mathcal B)$-correspondence $E = E(K)$ whose elements are $\mathcal B$-valued
        functions on $X \times \mathcal A$
        such that the function $k_{x} \colon (x',a') \mapsto
        K(x', x)[a']$ is in $E( K)$ for each $x \in X$
        and has the reproducing property
        \begin{equation*}  
     \langle k_{x}, a \cdot f  \rangle_{E(K)} =
     \langle  a^{*} \cdot k_{x} , f \rangle_{E(K)} =
     f(x,a)
     \text{ for all } x \in X \text{ and } a \in \mathcal A
        \end{equation*}
        where $a^{*} \cdot k_{x}$ is given by
        \begin{equation}  \label{concrete-Kol}
        (a^{*} \cdot k_{x})(x', a') = K(x', x)[a^{*}a'] = \langle a^* \cdot k_{x'} , a'*
        \cdot k_x \rangle .
        \end{equation}
        \end{enumerate}
        \end{theorem}

Corresponding
to the MVCS in (\ref{modCS}), define the {\em kernel} $K : X \times X \longrightarrow
\mathcal L ( \mathcal A , \mathcal B \otimes \mathcal C)$ by
\be
  K (x, y) a^* a^\prime = \langle x , a \mid y , a^\prime \rangle_\bH = \sum_k \langle
  a \cdot F_k (x) \otimes \phi_k \mid a^\prime \cdot F_k (y) \otimes \phi_k \rangle_\bH,
\label{mod-repker}
\en
 for all $ x,y \in X$  and $ a, a^\prime \in \mathcal A$ .
This is a completely positive kernel. In fact, (\ref{mod-repker}) gives the
 Kolmogorov decomposition of the kernel.

Let us see what reproducing property this kernel has.
\beano
k(x,z) a^* a^\prime & = & \langle x , a \mid z, a^\prime \rangle \\
& = & \langle x , a \mid I_\bH \mid z, a^\prime \rangle \\
& = & \langle x , a \mid \int_X \mid y,  b \rangle \langle y, b \mid \; d\mu(y)
\mid z, a^\prime \rangle \mbox{ for a co-isometry } b \in \mathcal A\\
 & = & \int_X \langle x , a \mid  y,  b \rangle \langle y, b  \mid z, a^\prime
 \rangle \; d\mu(y) \\
  & = & \int_X k( x , y) a^* b \; \; k( y, z)b^* a^\prime \; d\mu(y). \\
\label{mod-repker3}
\enano
In particular, taking $a = b = id_{\mathcal A}$, we get that $ k(x,z) a^\prime =
\int_X k( x , y) id_{\mathcal A} \; \; k( y, z)  \; d\mu(y) \; a^\prime$ which means
that $ k(x,z)  = \int_X k( x , y) id_{\mathcal A} \; \; k( y, z)  \; d\mu(y)$.

   We show the existence of an associated reproducing kernel and a carrier Hilbert module,
   corresponding to a family of MVCS.
Going back to the setting of Section \ref{sec:modCS}, we take again the measure space
$(X, \mu )$, the Hilbert modules
$\bE$ over the $C^*$-algebra $\mathcal B$, $\bG$ over the $C^*$-algebra $\mathcal C$,
$\bH = \bE\otimes\bG$ and the Hilbert module $\hh$, consisting of measurable functions
$F: X \longmapsto \bE$, which satisfy the ``square integrability condition'',
$$\Vert\int_X  \langle F(x) \mid F(x) \rangle_\bE
           \; d\mu (x) \Vert_{\mathcal B}< \infty\; . $$
We also require that the elements $\phi_i \in \bG$ used to define the non-normalized MVCS
in (\ref{modCS}) satisfy both the conditions (\ref{exframe}) and (\ref{orthonorm}).

   Consider now the Hilbert module $\widetilde{\hh}$, over $\mathcal B \otimes
   \mathcal C$,
consisting of maps $\widetilde{\bh} : X  \longrightarrow\mathcal B \otimes \mathcal C$,
under the (module) inner product,
$$
  \langle \widetilde{\bh}_1 \mid \widetilde{\bh}_2 \rangle_{\widetilde{\h}}
   = \int_X  \widetilde{\bh}_1 (x)^* \; \widetilde{\bh}_2 (x) \;  d\mu (x)\; . $$

   Recall that our coherent states $\vert x, a \rangle$ in (\ref{modCS}) are elements of
the Hilbert module $\bH$. Using these MVCS  we now define the linear map
$W: \bH \longmapsto \widetilde{\hh}$ by
\be
  (Wf)(x) =
    \langle x , \text{id}_{\mathcal A} \mid f \rangle_\bH   \; ,    \qquad f \in \bH\; .
\label{mod-isom1}
\en
That $W$ is an isometry is then clear, since
$$
 \langle Wf \mid Wf \rangle_{\widetilde{\h}} =
 \int_X \langle f \mid x,  \text{id}_{\mathcal A} \rangle\langle x ,  \text{id}_{\mathcal A}
 \mid f\rangle\; d\mu (x)
 = \langle f\mid f \rangle_\bH\; , $$
using (\ref{modresolid}).

\begin{theorem} The range of $W$ is a complemented submodule of
$\widetilde{\hh}$. \end{theorem}

\prf We denote by $\mathbb P_K$ the linear operator on
$\widetilde{\hh}$ defined by \be
  ( \mathbb P_K \widetilde{\bh} )(x) = \int_X K(x, y)\widetilde{\bh} (y)\;
  d\mu (y)\; , \qquad \mbox{for all~}
   \widetilde{\bh} \in \widetilde{\hh}\; .
\label{integ-ker} \en

It is then straightforward to verify that $\mathbb P_K$ is a
projection in the $C^*$-algebra $\mathcal L ( \widetilde{\hh} )$
and the range of the isometry in $\widetilde{\hh}$, which we
denote by $\hh_K$, is range of the projection $\mathbb P_K$.
The range of a projection is always a complemented submodule. \qed

We call it a {\em reproducing kernel submodule} because it
is the image under an isometry of an $\mathcal A$-$\mathcal B$
correspondence. The reproducing kernel $K(x, y)$ is $\langle x, \text{id}_{\mathcal A}
\mid y, \text{id}_{\mathcal A}\rangle$. It follows that
$$\widetilde{\hh} = \hh_K \oplus \hh_K^\perp.$$

  Writing
\be
   \bh_x = W\mid x, \text{id}_\mathcal A\rangle \in \widetilde{\hh}\; , \quad \text{so that}
   \quad K(x, y) = \langle \bh_x \mid \bh_y\rangle_{\widetilde{\h}} = \bh_y (x)
   \in \mathcal B
   \otimes \mathcal C\; ,
\label{MVCS-2}
\en
we see that the vectors $\bh_x$ span the submodule $\hh_K$. From
(\ref{modresolid}), (\ref{integ-ker})  and (\ref{MVCS-2}), it also follows that
\be
   \int_X \mid \bh_x \rangle \langle \bh_x \mid\; d\mu (x) = \mathbb P_K\; .
\label{MVCS-resolid4}
\en
Note that the vectors $\bh_x , \; x \in X$, being unitary images in $\hh_K$ of the MVCS
$\mid x, \text{id}_\mathcal A \rangle$, are also themselves MVCS.
Furthermore, the submodule $\hh_K$ has a natural left action for $a \in \mathcal A$ given by
$$ (a \cdot \widetilde{\bh})(x) = (Waf)(x)\; , \quad \text{where}\quad \widetilde{\bh}
= Wf\; .  $$

  Finally, using the $\bh_x$ we may define a POV measure on $\hh_K$ and obtain a natural
  dilation of it to a PV measure on $\widetilde{\hh}$. Indeed, the POV measure is defined
  on the Borel sets $\Delta$ of $X$ as,
\be
   \nu(\Delta ) = \int_\Delta \mid \bh_x \rangle \langle \bh_x \mid\; d\mu (x) \in
   \mathcal L (\hh_K)\; ,
\label{pov-meas}
\en
and the PV measure $\widetilde{P}(\Delta )$ by
\be
 ( \widetilde{P}(\Delta )\widetilde{\bh} )(x) = \chi_\Delta (x )\widetilde{\bh} (x)\; ,
 \qquad \widetilde{\bh} \in \widetilde{\hh}\; ,
\label{PV-meas}
\en
$\chi_\Delta$ being the characteristic function of the set $\Delta$. It is then
straightforward
to verify that
\be
   \nu (\Delta ) = \mathbb P_K \widetilde{P}(\Delta)\mathbb P_K\; .
\label{dilation-prop}
\en
If $X$ is a locally compact space and the support of the measure $\mu$  is assumed to
be the whole of $X$ (i.e., no open set has measure zero), this dilation can easily be
shown to be minimal, in the sense of Naimark.  In other words,  the
set of vectors of the type
$\widetilde{P}(\Delta )\bh$, as $\Delta$ runs through all Borel sets and $\bh$ through
$\hh_K$, spans $\widetilde{\hh}$. The proof is an easy adaptation of the proof of the analogous result for
Hilbert spaces (see, e.g., \cite[p. 36]{aag-book}).

  The space $\widetilde{\hh}$ acts as a carrier space for the MVCS, the situation
with the dilation here being exactly the same as on a Hilbert space. It would appear that
the fundamental ingredients  for the existence of a family of MVCS are ($i$) a
Hilbert $C^*$-module of
the type $\widetilde{\hh}$, consisting of functions from a finite measure space $(X, \mu )$ to
the $C^*$-algebra defining the module and $(ii)$ a reproducing kernel submodule contained in
this Hilbert module. We plan to discuss these issues in greater
detail in a succeeding publication.

\section{Appendix}\label{sec:appendix}

\subsection{Hilbert $C^*$-modules}\label{subsec:hilbmod}
In this appendix we collect together some preliminary notions and results on Hilbert modules.
Let $A$ be a $C^*$-algebra (not necessarily unital or commutative).
An {\it inner-product $A$-module} is a linear space $E$ which is a right
$A$-module ( with compatible scalar multiplication: $\lambda(xa)=(\lambda x)a=
x (\lambda a)$ for $x\in E, a\in A, \lambda\in \mb C$), together with a map
$(x,y)\mapsto \langle x\mid y\rangle$ from $E\times E\longrightarrow A$ such that
\begin{enumerate}
\item[(i)]$ \langle x\mid \alpha y+\beta z\rangle=\alpha\langle x\mid y\rangle+
\beta\langle x\mid z\rangle$\hspace*{50pt} $(x,y,z\in E,\alpha,\beta\in \mb C),$
\item[(ii)] $\langle x\mid ya\rangle=\langle x\mid y \rangle a$ \hspace*{130pt}
$(x,y\in E, a\in A),$
\item[(iii)] $\langle y\mid x \rangle=\langle x\mid y \rangle^* $ \hspace*{135pt}
$(x,y\in E),$
\item[(iv)] $\langle x\mid x \rangle\geq0;$ \hspace*{17pt} if $\langle x\mid x
\rangle=0$ then $x=0.$
\end{enumerate}
Note that in condition (i) the inner-product is required to be linear in
its {\it second} variable. From (iii) it follows that the  the inner-product is
conjugate-linear in its { first} variable. We adopt the same convention for ordinary
inner-product spaces and Hilbert spaces (so that an inner-product space is the same
thing as an inner-product $\mb C$-module). If $E$ satisfies all the conditions for
an inner-product $A$-module except for the second part of condition (iv) then we
call $E$ a {\it semi-inner-product $A$-module}.  For such modules we have a
version of Cauchy-Schwarz inequality:
\begin{prop}\cite[p. 3]{lan}\label{CS}
If $E$ is a semi-inner-product $A$-module and $x,y\in E$ then
$$\langle y\mid x \rangle\langle x\mid y \rangle\leq \Vert\langle x\mid x
\rangle\Vert \langle y\mid y \rangle$$
\end{prop}

For $x\in E$ we write $\Vert x\Vert=\Vert\langle x\mid x \rangle\Vert^\frac{1}{2}.$
It follows from Proposition \ref{CS} that $ \Vert \langle x\mid y \rangle\Vert\leq
 \Vert x\Vert\Vert y\Vert$ and it is easy to deduce from this that if $E$ is an
 inner-product $A$-module then $\Vert .\Vert$ is a norm on $E.$ An inner-product
 $A$-module which is complete with respect to its norm is called a {\it Hilbert
  $A$-module}, or a {\it Hilbert $C^*$-module over the $C^*$-algebra $A$.}

One would like to think that Hilbert $C^*$-modules behave like Hilbert spaces,
and in some ways they do. For example, if $E$ is a Hilbert $A$-module and $x\in E$
then it is easy to check that
$$\Vert x\Vert=\mbox{~sup~}\{\Vert\langle x\mid y \rangle\Vert : y\in E,
\Vert y\Vert\leq 1\}.$$
But there is a fundamental way in which Hilbert $C^*$-modules differ from Hilbert
spaces. Given a closed submodule $F$ of a Hilbert $A$-module $E$, define
$$F^\perp=\{y\in E: \langle x\mid y \rangle=0, (x\in E)\}$$
Then $F^\perp$ is also a closed submodule of $E$. But $E$ is not (usually) equal to
$F\oplus F^\perp$ (and $F^{\perp\perp}$ is usually bigger than $F$) \cite[p. 7]{lan}.

Suppose that $E,F$ are Hilbert $A$-modules. We define $\mathcal L(E,F)$ to be the set
of all linear($\mb C$-linear) maps $t:E\longrightarrow F$  for which there is a
linear map $t^*:F\longrightarrow E$
such that
 $$\langle tx\mid y\rangle=\langle x\mid t^*y\rangle \hspace*{40pt}(x\in E, y\in F).$$
 It is easy to see that $t$ must be $A$-linear (that is, $t$ is linear and $t(xa)=t(x)a$
 for all $x\in E, a\in A$) and bounded as a map between the Banach spaces $E$ and $F$.
  We call $\mathcal L(E,F)$ the set of {\it adjointable} maps from $E$ and $F$. In
  particular, $\mathcal L(E,E)$ which we abbreviate to $\mathcal L(E),$
  is a $C^*$-algebra. Thus every element of $\mathcal L(E,F)$ is a bounded $A$-linear map.
  But the converse is false:  a bounded $A$-linear map need not be
  adjointable \cite[p. 8]{lan}.

We say that a closed submodule $F$ of a Hilbert $A$-module is {\it complemented}
if $F\oplus F^\perp.$ As already emphasized that a closed submodule of a Hilbert
$C^*$-module need not be complemented. The following theorem enables us to conclude
that certain submodules are complemented.
\begin{thm}\cite[p. 22]{lan}
Let $E,F$ be Hilbert $A$-modules and suppose that $t$ in $\mathcal L(E,F)$ has
closed range. Then
\begin{enumerate}
\item[(i)] {\rm ker}$(t)$ is a complemented submodule of $E$.
\item[(ii)] {\rm ran}$(t)$ is a complemented submodule of $F$.
\item[(iii)] the mapping $t^*\in \mathcal L(E,F)$ also has closed range.
\end{enumerate}
\end{thm}
An operator $u\in \mathcal L(E,F)$ is said to be a {\it unitary} if $u^*u=1_E$
and $uu^*=1_F$. If there exists a unitary element of $\mathcal L(E,F)$ then we
say that $E$ and $F$ are {\it unitarily equivalent} Hilbert $A$-modules, and we
write $E\approx F.$ The following two results characterise  unitary maps and
isometries from Hilbert $A$-modules $E$ to $F$ respectively.
\begin{thm}\cite[p. 26]{lan}
Let $A$ be a $C^*$-algebra, let $E,F$ be Hilbert $A$-modules and $u$ be a linear
map from $E$ to $F$. Then the following conditions are equivalent:
\begin{enumerate}
\item[(i)] $u$ is an isometric, surjective $A$-linear map;
\item[(ii)] $u$ is a unitary element of $\mathcal L(E,F)$.
\end{enumerate}
\end{thm}
\begin{prop}
With $A,E,F$ as before, let $w$ be a linear map from $E$ to $F$. The following
conditions are equivalent:
\begin{enumerate}
\item[(i)] $w$ is an isometric $A$-linear map with complemented range;
\item[(ii)]$w\in \mathcal L(E,F)$ and $w^*w=1_E.$
\end{enumerate}
\end{prop}
Let $A,B$ be $C^*$-algebras and let $E$ be a Hilbert $B$-module. Suppose that
$\phi:A\longrightarrow \mathcal L(E)$ is a $*$-homomorphism of $C^*$-algebras.
We can also regard $E$ as a left $A$-module, the action being given by
$$(a,y)\mapsto\phi(a)y\hspace*{40pt} (a\in A, y\in E).$$
In this situation, we call $E$ is a $A-B$ correspondence.

If $\mathcal H$ is a Hilbert space then the algebraic (vector space)
tensor product $\mathcal H\otimes_{\mbox{alg}}A$ (which is a right $A$-module,
the module action being $(\xi\otimes a)b=\xi\otimes ab\hspace*{5pt}(\xi\in\mathcal H;
a,b\in A)$) has an $A$-valued inner-product given on simple tensors by
$$\langle \xi\otimes a\mid \eta \otimes b\rangle=\langle \xi\mid \eta \rangle
 a^*b\hspace*{50pt} (\xi,\eta\in \mathcal H, a,b\in A).$$ \
It can be verified that this is a positive definite inner-product on
 $\mathcal H\otimes_{\mbox{alg}}A$ \cite[p. 6]{lan}. Thus $\mathcal
 H\otimes_{\mbox{alg}}A$ is an inner-product $A$-module, we denote its
 completion by $\mathcal H\otimes A.$ In the case where $\mathcal H$ is a
 separable, infinite-dimensional Hilbert space, the Hilbert $A$-module $\mathcal H\otimes A$ is often denoted by $\mathcal H_A.$ If $E$ is a Hilbert $A$-module and $Z\subseteq E$ then we say that $Z$ is a {\it generating set} for $E$ if the closed submodule of $E$ generated by $Z$ is the whole of $E$.  We say that $E$ is countably generated if it has a countable generating set. We now state a theorem, known as Kasparov's stabilisation theorem. Intuitively, the idea of the theorem is that $\mathcal H_A$ is big enough to absorb any countably generated Hilbert $A$-module; or alternatively that once a module reaches the size of $\mathcal H_A$ it stabilises, cannot get any "bigger".
\begin{thm}\cite[p. 60]{lan}
If $A$ is a $C^*$-algebra and $E$ is a countably generated Hilbert $A$-module then $E\oplus\mathcal H_A\approx\mathcal H_A.$
\end{thm}
We say that a closed submodule $F$ of a Hilbert $C^*$-module $E$ is {\it fully complemented} if $F$ is complemented in $E$ and $F^\perp\approx E.$
\begin{cor}
If $E$ is a countably generated Hilbert $A$-module then $E$ is unitarily equivalent to a fully complemented submodule of $\mathcal H_A.$
\end{cor}
Suppose that $A,B$ are $C^*$-algebras, $E$ is a Hilbert $A$-module and $F$ is a Hilbert $B$-module. We want to define $E\otimes F$ as a Hilbert $(A\otimes B)$-module. Start by forming the algebraic tensor product $E\otimes_{\mbox{alg}}F$ of the vector spaces $E$ and $F$ (over $\mb C).$ This is a right module over $A\otimes_{\mbox{alg}}B$ (the module action being given by $(x\otimes y)(a\otimes b)=xa\otimes yb$). For $x_1, x_2$ in $E$ and $y_1, y_2$ in $F$, we define
$$\langle x_1\otimes y_1\mid  x_2\otimes y_2\rangle=\langle x_1\mid x_2\rangle \otimes \langle y_1\mid y_2\rangle.$$
This extends by linearity to an  $(A\otimes_{\mbox{alg}}B)$-valued sesquilinear form on  $E\otimes_{\mbox{alg}}F$ which makes  $E\otimes_{\mbox{alg}}F$ into a semi-inner-product $( A\otimes_{\mbox{alg}}B)$-module over the pre-$C^*$-algebra  $A\otimes_{\mbox{alg}}B$ \cite[p. 34]{lan}. It can be shown that the semi-inner-product on  $E\otimes_{\mbox{alg}}F$  is actually an inner-product  by Kasparov's stabilisation theorem \cite[p. 62]{lan}. The double completion process \cite[p. 4]{lan} can be performed to conclude that the completion $E\otimes F$ of on  $E\otimes_{\mbox{alg}}F$ is a Hilbert $(A\otimes B)$-module. We call $E\otimes F$ the {\it exterior tensor product} of $E$ and $F$.
\begin{prop}
With notations as in  in Section \ref{sec:modCS}. We have $\hh\approx L^2(X,\mu)\otimes E.$
\end{prop}
\prf  The unitary map $v: L^2(X,\mu)\otimes E\longrightarrow \hh$ is given by defining  $v(F\otimes\xi)$ to be the map $x\mapsto F(x)\otimes\xi ~~(x\in X) $, where   $F\in L^2(X,\mu), \xi\in E.$ It is straightforward to verify that $v$ is a unitary.

\subsection{Bijection}\label{bij}
Here we write down an explicit bijection from $\mathbb N$ to $\mathbb N\times \mathbb N.$
\begin{prop}
For $k \in \mathbb N,k>1$ we have
 \begin{enumerate}
 \item[(i)] there is a unique integer $n_k \in \mathbb N$ such that $\frac{n_k(n_k+1)}{2}< k \leq \frac{(n_k+1)(n_k+2)}{2}$

 \item[(ii)]$t(k):= \left\{
                      \begin{array}{ll}
                        (1,1), & \hbox{for $k=1$;} \\
                        \big(\frac{n_k(n_k+3)}{2}+2-k,k-\frac{n_k(n_k+1)}{2}\big), & \hbox{for  $k>1$.}
                      \end{array}
                    \right.$

 is a one-to-one map from $\mathbb N$ onto $\mathbb N\times \mathbb N.$
\end{enumerate}
\end{prop}
\prf We observe that $\mathcal I:=\{\big(\frac{n(n+1)}{2},\frac{(n+1)(n+2)}{2}]: n\in \mb N\}$ is a family of non-intersecting intervals in the real line whose union is the interval $(1,\infty).$ So any real number $x>1$ can belong to a unique interval in $\mathcal I$. In particular, this is true for any  $k\in\mb N,k>1$. This proves $(i).$

To prove $t$ is one-to-one we show that $t(k)=t(k^\prime)$ implies $k=k^\prime$.
If $n_k=n_k^\prime$ then it is clear from the expression for $t(k)$ that $t(k)=t(k^\prime)$ implies $k=k^\prime$. To settle the other possibility, let $n_k\neq n_k^\prime$ and $t(k)=t(k^\prime)$.  By hypothesis we have $\big(\frac{n_k(n_k+3)}{2}+2-k,k-\frac{n_k(n_k+1)}{2}\big)=\big(\frac{n_{k^\prime}(n_{k^\prime}+3)}{2}+2-k^\prime,k^\prime-\frac{n_{k^\prime}(n_{k^\prime}+1)}{2}\big).$
Equating the first component of the ordered pairs we have $k^\prime-k=\frac{1}{2}(n_{k^\prime}-n_k)(n_{k^\prime}+n_{k}+3)$. Similarly, we obtain $k^\prime-k=\frac{1}{2}(n_{k^\prime}-n_k)(n_{k^\prime}+n_{k}+1)$ from the second component of the ordered pair. Equating the expressions for $k^\prime-k$ we get $n_k=n_{k^\prime}$.  Therefore, we are reduced to the previous case and hence $k=k^\prime$.

Finally, we show that show that $t$ is onto. Given any $(p,q)\in\mb N\times\mb N$ we  set $n_k=p+q-2$ and $k=\frac{n_k(n_k+3)}{2}+2-p=\frac{n_k(n_k+1)}{2}+q$.  We see that $t(k)=(p,q).$ This completes the proof.


\end{document}